# Finlay, Thames, Dufay, and Paget color screen process collections: Using digital registration of viewing screens to reveal original color

**Geoffrey Barker [1], Jan Hubička[2], Mark Jacobs [3], Linda Kimrová[4],**
**Kendra Meyer[5], Doug Peterson[6]**

[1]State Library of New South Wales

[2]Department of Applied Mathematics, Charles University; Šechtl and Voseček Museum of Photography; SUSE LINUX s.r.o.

[3]In memoriam

[4]Charles University

[5]American Museum of Natural History

[6]DT Heritage

Contact: Jan Hubička, hubicka@kam.mff.cuni.cz

**Abstract**
We discuss digitization, subsequent digital analysis and processing of negatives (and diapositives) made by Finlay, Thames, Dufay, Paget, and similar additive color screen processes. These early color processes (introduced in the 1890s and popular until the 1950s) used a special color screen filter and a monochromatic negative. Due to poor stability of dyes used to produce color screens many of the photographs appear faded; others exist only in the form of (monochromatic) negatives. We discuss the possibility of digitally reconstructing the original color from scans of original negatives or by virtue of infrared imaging of original transparencies (which eliminates the physically coupled color filters) and digitally recreating the original color filter pattern using a new open-source software tool. Photographs taken using additive color screen processes are some of the very earliest color images of our shared cultural heritage. They depict people, places, and events for which there are no other surviving color images. We hope that our new software tool can bring these images back to life.

**Introduction**
Although the first attempts to produce color photography date back to the late 1840s (Edmond Becquerel), the principle of color photography as we know it today was introduced/demonstrated by James Clerk Maxwell during a lecture before the Royal Institution in London on May 17, 1861. Maxwell asked the photographer and inventor Thomas Sutton to take three separate black and white negatives through three colored liquid filters (red, green, and blue) which resulted in three *separation negatives*. These negatives were copied to positive transparencies (*separation transparencies*) and projected through filters of the same color to demonstrate practicality of the additive color synthesis principle. Due to lack of sensitivity of photographic emulsions to red light this experiment was only partly successful. However, this principle was later developed into a practical method of *three-color photography*. One of the main drawbacks was the difficulty involved in viewing photographs: it was either possible to use expensive projectors or chromoscopes to combine 3 separation transparencies into a full color image or print the photographs using very laborious method of color dye-transfer processes. Consequently, color photographs were not widely available and known. See also (Lavédrine and Gandolfo, 2013) and (Pénichon, 2013).

To simplify the process Louis Ducos du Hauron, in 1869, proposed an idea of color photography process which used only one (monochromatic) negative. Following his idea John Joly, in 1884, and James William McDonough, in 1896 with first patent in 1892, introduced the first commercially available implementations of this method: While taking a photograph, a special *taking screen* consisting of regular red, green and blue lines (each about 0.1mm wide) was placed directly on the top of orthochromatic negative inside of camera. Later the negative was copied to a positive monochromatic transparency. Finally, the original color was reproduced by registering the transparency with a *viewing screen* consisting of the same color pattern.

**Joly Colour screens** were produced from 1896–1900, **McDonough plates** 1896–1900, but their practical use was limited by long exposure times and lack of panchromatic negatives. The idea was developed further with improvements in exposure time, color quality and applied by several commercial color processes (Casella and Cole, 2022): **Thames Colour Screen** (1908–1910), **Dufay Dioptichrome** (1909–1910), **Paget Color screen** (1913 – ca. 1922), **Duplex Screen Plate** (1926 – ca. 1928), **Finlay Colour Plate** (1929–1941, Fig. 1), **Johnsons Colour Screens** (1953 – ca. 1954).

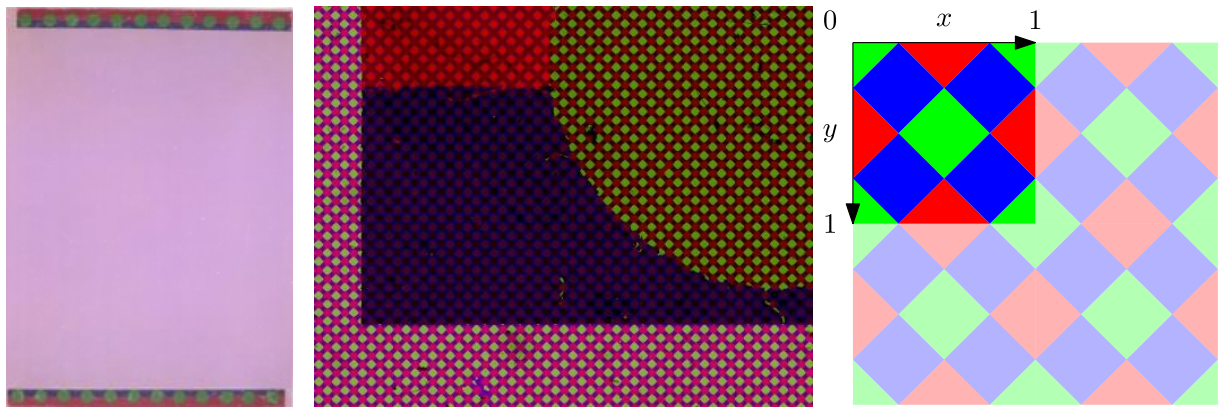

Fig. 1 - Finlay taking screen 13x18cm (left), a detail of a 5400PPI scan using Eversmart Supreme (middle) and a digital model (right). These screens are characteristic by registration marks (a strip with green disks) along top and bottom edges of the plate simplifying the viewing screen registration.

While these processes differed in many details, they all share the key feature of using taking screen in combination with an (orthochromatic, later panchromatic) black and white negative. In all cases the transparency is later registered with a viewing screen to produce a color image.

Several other processes employ the same basic principle, however making the color screen (or *réseau*) integral part of the glass plate or film containing the emulsion. Known examples include (Casella and Cole, 2022): **Thames Colour Plate** (1909–1934), **Omnicolore** (1909–1911), **Dufay Dioptichrome-B Plate** (1910–1912), **Dufay Improved Dioptichrome-B Plate** (1912–1914), **Finlaychrome** (ca. 1933–1940), **Krayn Line Screen** (1909–1911), **Krayn Color Film** (1910–1911), **Dufaycolor Film** (1935–1958), and **PolaChrome** (1983–2002).

All those processes are examples of *additive color screen processes* (with regular screens as opposed to autochrome and Agfa color screen plates which uses *random* or *stochastic screen*) on which techniques discussed in this paper apply. In most cases the color filter physically attached to the negative has faded or physically warped in a way that totally obfuscates the color originally recorded of the scene. However, the unique geometrically regular pattern of these processes allows us to digitally reconstruct the original scene color. This is not attempting to shift or adjust the image based on the color as seen today, but a mathematical reconstruction of that color based on a first principles understanding of what each color in these processes was at the time of capture.

**Producing an additive color screen photograph**
To take a color photograph using an additive color screen filter one proceeds in following basic steps:
**1.** Fix taking screen filter to the emulsion side of an ordinary (ideally panchromatic) black and white negative. (It is important that the color screen is in direct contact with the emulsion and thu binding tape was used to fix glass plates together.) **2.** Expose the photograph the same way as when taking black and white photographs (with a longer exposure time accounting for the filter.) **3.** Develop the black and white image. **4.** Use contact copying to produce a black and white transparency. **5.** Register the viewing screen filter to the transparency to reconstruct colors.

Steps 2–4 are quite standard, and every photographer of that era could do them. The last step, *registration of the viewing screen*, involves determining rotation and translation of the viewing screen to match the original position of the taking screen and binding it to the transparency using binding tape for magic lantern slides. This needs a great degree of precision, since color patches are small (1/10$^{th}$ of millimeter in the case of Paget and Finlay screen processes). Today it is common that the transparencies completely or partly lost registration resulting in a color quality loss (Fig. 2).

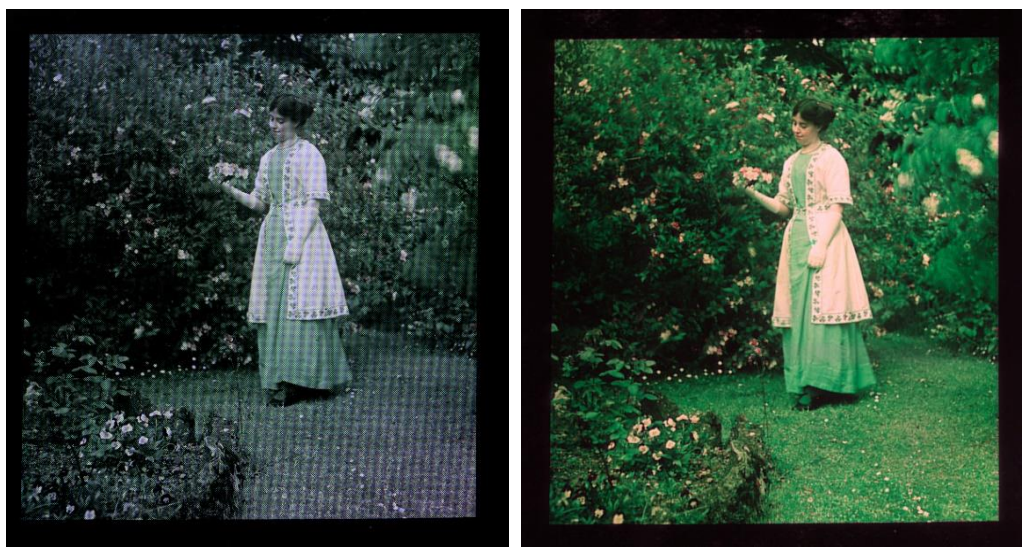

Fig. 2 - Paget color transparency 8×8cm (left), ca. 1921, author unknown, Šechtl and Voseček Museum of Photography. Probably due to aging of the binding tape, the viewing screen is rotated by approximately 0.04° resulting in loss of color in the upper left and lower right part of the image. Digital color reconstruction based on an infrared scan of the transparency (right) eliminates the problem. Note that by studying the dot pattern of the transparency one can verify that the image originally indeed had a strong green-yellow cast.

**Identification of negatives taken using additive processes with non-integrated screens**
Negatives taken using regular additive color processes can be easily mistaken for normal black and white photographs, since the pattern originating from the color screen is visible only at magnification (or in high PPI scan). It is also quite easy to mistake the regular pattern (recording the color information) for a reproduction of a halftone print (Meyer, 2020). See Fig 3.

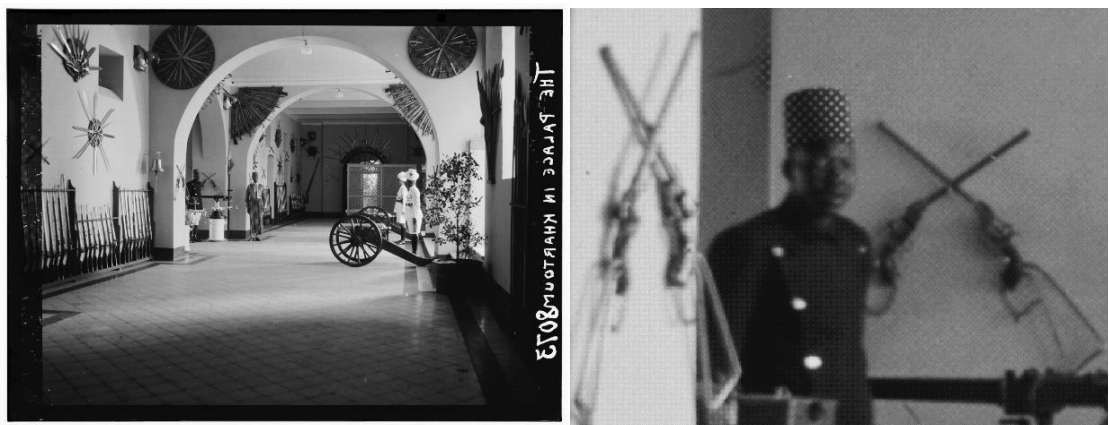

Fig. 3 - Matson Photo Service: Sudan. Khartoum. Hallway of the Palace, cannon facing entrance, 1836.
Library of Congress, Prints and Photographs Division LC-DIG-matpc-00295.
A negative taken using Finlay Color screen digitally inverted to positive (left) and detail of scan at 1000PPI showing regular pattern well visible in bright colors. One can also notice barely visible registration marks (compare with Fig. 1) along the edges of the negative.

*Colour Photography and Film:*
*sharing knowledge of analysis, preservation, conservation, migration of analogue and digital materials*

Presently we know of the following collections which contains substantial number (over 10) of photographs taken using regular color screen processes:

1. *[Matson (G. Eric and Edith) Photograph Collection (the American Colony collection)](#)* at the Library of Congress. This collection contains 179 Finlay color negatives and 3 color transparencies. Most photographs were taken in Africa in 1936. Earliest (taken in Betlehem) are dated 1934 and latest (taken in the archaeological site of Petra) are dated 1946.

2. *[130 Paget plate negatives by Yvette Borup Andrews](#)* taken during the first Asiatic Zoological Expedition 1916–1917 in collection the American Museum of Natural History.

3. *[Paget plates (positives with viewing screens attached) by Frank Hurley](#)* from the British Imperial Trans-Antarctic Expedition, 1914–1917, 32 photographs and ca. 30 Paget plates (also positives) by Frank Hurley taken during WW1 in the collection of the State Library of New South Wales.

4. *[Paget plate transparencies with viewing screens attached and negatives by Frank Hurley](#)* taken during WW1 in the collection of the Australian War memorial. Ca. 300 photographs.

5. *[Finlay color negatives and transparencies by Oscar Jordan](#)* at the Franklin D. Roosevelt Presidential Library and the Herbert Hoover Presidential Library.

6. **The National Geographic collection** contains several Paget, Finlay and Dufaycolor transparencies. This collection is slated for mass digitization at FADGI 4 standards by Digital Transitions / Pixel Acuity in late 2022.

7. *[Collection of approximately 320 lantern slides made by the Paget Colour process by Henden Hardwick](#)*, depicting views in London, Hampstead and the Wye Valley listed at Science Museum Group catalogue. The collection contains also lecture notes and negatives.

8. **Private collection of Mark Jacobs**, while focused on Autochrome, contains examples of transparencies by Joly, Paget, Finlay, Dioptichrome and Dufay color screens.

9. **Collection of color photographs taken by Oscar Jordan at the Franklin D. Roosevelt Presidential Library and Museum**. This collection contains probably hundreds of large format (8x10in) negatives and transparencies taken using Finlay color screens.

Collections of negatives are hard to find. We believe that many additional examples remain hidden and have yet to be identified as color photographs. We hope that our work will help this process.

**Digital processing of additive color screen photographs (with regular screens)**
We started by implementing simple GUI to aid digital registration of the viewing screen to the scan of the negative which developed to a specialized open-source tool (*[ColorScreen](#)*). Presently the tool consists of a queue of image transformations which resembles ones performed by tools for processing raw image data from digital cameras (since modern digital cameras use a Bayer filter which is an additive color screen process):

> scan linearization & channel mixing → sharpening → simulation of negative-to-positive process → viewing screen registration → simulation of viewing screen → demosaicing → recovery of fine details → simulation of color dyes in the viewing screen → brightness, white balance and saturation adjustments

We briefly outline the four steps most specific to processing early color photography.

*Viewing screen registration*: This step establishes a map between scan-coordinates and screen-coordinates. Scan-coordinates address pixels in the input image. Screen-coordinates are defined in a way so the whole screen can be produced by repeating a tile of size 1×1 (see Fig. 1, right).

We found this step to be difficult (impossible to do in general purpose image manipulation programs) since scans are not produced by contact copying and it is often necessary to compensate for several geometric errors. In addition to rotation and translation we also compensate for scaling (caused by slight deviation of the photographic emulsion from scanning plane), skewing (caused by sensor rotation in scanner with linear sensors), scanner lens distortion and perspective correction (since the original may be slightly tilted from the scanning plane). All these corrections were motivated by real problems in the scans. Many scans contain additional errors, most notably caused by imprecisions of the stepping motor moving the sensor for which we have not yet compensated.

*Simulation of Viewing screen*: Now it is possible to simulate the viewing screen filter placed on the top of the transparency; for every pixel of the input image the RGB color of the viewing screen is determined and these RGB values are multiplied by the luminosity of the pixel.

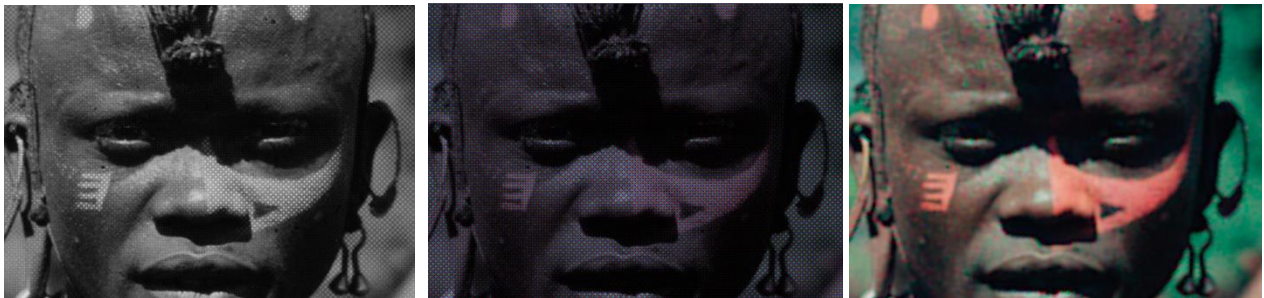

Fig. 5 - Original scan (left), result of viewing screen simulation (middle) and demosaiced image using bicubic interpolation (right).

*Demosaicing (optional step):* While we now can closely approximate the actual additive color transparency, it is difficult to display or print the results since color screen processes relies on high resolution, high dynamic range, and saturated colors (see Fig. 5, left). Because colors in screens are organized in an analogous way as colors in the Bayer filter used in modern camera, it is possible to compute luminosity of each patch (by collecting weighted average of scan pixels mapped to it) and apply standard demosaicing algorithms (such as bicubic interpolation) to obtain a smooth picture with full RGB information for each pixel (see Fig. 5, right). This picture is however not very sharp: 8×8cm Paget plate yields to digital equivalent of 0.6-megapixel photograph which we compensate for later by another optional step with a specialized algorithm.

*Simulation of viewing screen dye color:* In this step we turn RGB values in the "color profile of the original process" computed earlier into RGB or XYZ values of the output color space. Towards this we collected published measurements of the spectra of dyes used by some of the additive color processes (Trumpy *et al.*, 2021) and simulate their color when viewed by the CIE 1931 Standard Observer under simulated daylight. Where available we can also simulate the color based on a spectral measurement of reconstructed filters (Lavédrine, 1992), (Wagner, 2006), (Casella and Tsukada, 2012). However, dyes used by individual color screens yet need to be understood.

**Digitization guidelines**
Precise analysis of the original transparency or negative is possible only when the digital scan meets certain minimum criteria. It is necessary to ensure the following:

*Resolution*: Since the color screen processes are based on contact copying of the taking screen to the photographic negative their resolution exceeds the resolution of ordinary black and white images. Minimal acceptable resolution depends on the size of color patches in the color screen. It is necessary to set resolution, so that each color patch is at least 2 pixels in diameter (this is approximately 500PPI for Paget and Finlay screens). This is by Nyquist theorem also the minimal resolution. The resulting color quality of our reconstructions noticeably increases until 8 pixels in diameter is met (approximately 2000PPI for Paget and Finlay screens). Where possible, even higher resolution can provide greater confidence in the result.

*Sharpness*: The optical sharpness of the scans (spatial frequency response, or SFR) should be good enough to render clearly the viewing screen. This is a problem especially for finer color screens (such as Dufaycolors) where at least 3000PPI scans are necessary. We recommend following FADGI 4-star guidelines for SFR 10, SFR50. However, we suggest keeping original files before any digital sharpening is applied. While not implemented yet, sharpening patterns created by color screen differs from usual sharpening and we expect specialized algorithms to lead to better results.

*Geometric precision*: Since taking screens are quite fine, significantly higher geometric precision is required than for digitization of standard photographic material. It is necessary to test individual scanners and understand their errors and see if they can be compensated for. We recommend producing a detailed documentation of the digitizing setup which included details such as type of device used (single-shot digital camera, digital camera with moving sensor or scanner with moving lenses), lenses used, focus distance, and other details which may affect the geometry of the scans.

*Preserving relative densities*: To obtain realistic results it is necessary to keep a connection between the densities of the transparency (or negative) scanned and the digital data stored in the scan. This is possible with raw scans but any digital image enhancement (such as dynamic range compression, sharpening, noise removal) is harmful. Where available we recommend saving linear TIFF files in the original profile of the device used to capture the image. A reference scan of an imaging target with known density such as the DT Trans OLT Target in the same capture frame or made in a separate capture with the same settings, can provide further documentation of the density at the time of capture. A proper even-fielding of the capture is also necessary.

*Dynamic range*: We found that negatives for additive color screen processes are quite contrasty. If the scanner is not able to digitize the full dynamic range of the original, the color information will be damaged or fully lost in highlights, shadows, or very saturated parts of the image.

*Infrared scanning:* Dyes used to manufacture color screens are transparent in infrared spectra. This makes it possible to use infrared scanning to obtain scan of the original monochromatic transparency which is sandwiched with the color screen and reconstruct original colors of transparencies which faded or got misregistered. Some scanners provide infrared scanning to aid dust and scratches removal of subtractive color films.

*Light and heat considerations*: While scanners are believed to be safe for digitizing standard color materials, extra consideration should be given when scanning color screens, especially in the infrared spectra. The process required development of very thin filters which easily peel when heated. In fact, while working on the project we damaged one Finlay taking screen which peeled during an experiment which involved about 1 hour of scanning at different settings and resolutions.

We tested scans made with the following devices:

*Epson Perfection V850 scanner*. This scanner features an infrared lamp, and it is possible to save linear TIFF files using the third-party VueScan software. The optical resolution about 2000PPI-3000PPI is good (but not ideal) for Paget and Finlay screens. It is not good for finer screens, such as Dufaycolors. To our surprise, we found that the dynamic range is not sufficient for significant percentage of Paget plate negatives from the Yevette Borup collection (while it works well for regular glass plate negatives). Finally, we fear that the scanner may be dangerous for scanning materials that involve color screens in infrared due to the heat it produces during slow high-resolution scans.

***Nikon Coolscan 9000ED***: This scanner has optical resolution close to 4000PPI which is enough to capture screens of Dufaycolor slides. It is limited to 6×9cm originals and shows noticeable stepping motor errors which produce a banding in scans. The depth of focus is small, and it is hard to get sharp scans even with glass holders. This device can produce perfectly aligned RGB and infrared scans, making it easy to analyze Dufaycolor color screens and produce quality renderings.

***DT Atom Rainbow MSI Station*** (with Eureka narrowband multispectral lights and the broad-spectrum high-CRI DT Photon white light source, a Schneider 120mm ASPH lens and a DT Phase One iXH 150mp camera)**:** This setup can achieve enough of optical resolution and dynamic range to obtain very good captures of Paget and Finlay color screen negatives. In addition, the provided software supports export in "linear scientific" files which precisely preserve relative color densities needed to get the colors right. DT Photon light source is 37° C and roughly room temperature at the point the film is held which, together with very short scanning times, eliminates the fear of light and temperature damage. Because infrared imaging is possible, it is also ideal for photographing transparencies. While not tested, we believe it to be an excellent solution for other additive screen processes as well, including Dufaycolor.

**Examples**

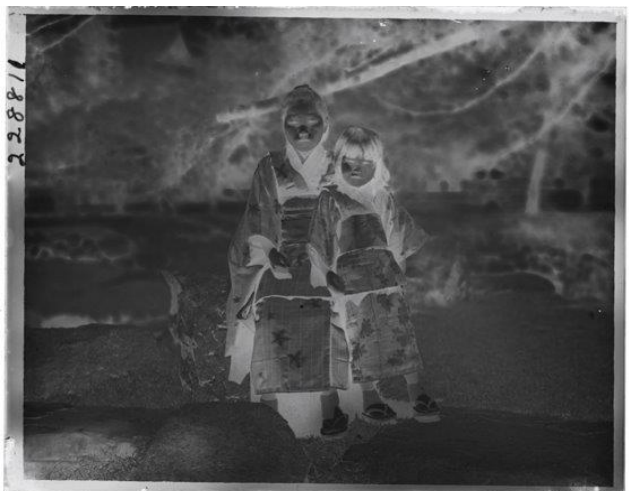 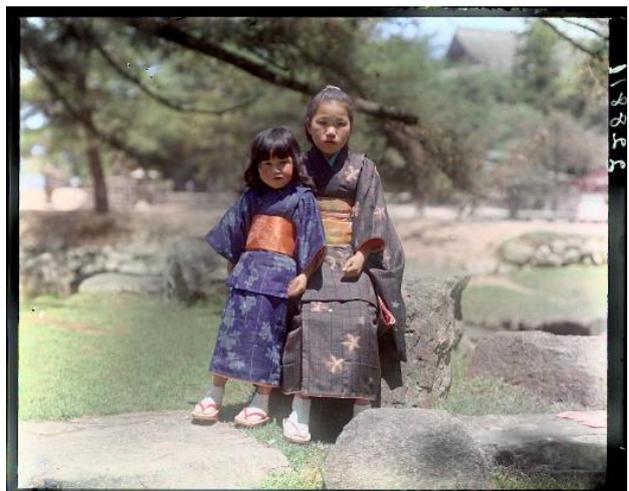

Fig. 6 - Yvete Borup Andrews: Two Japanese Children, Kyoto, First Asiatic Expedition. April 1916. Paget plate negative 4×5in, AMNH Special Collection, 228811 (left) and digital color reconstruction based on the negative. Scanned at 2400PPI using Epson Perfection V850.

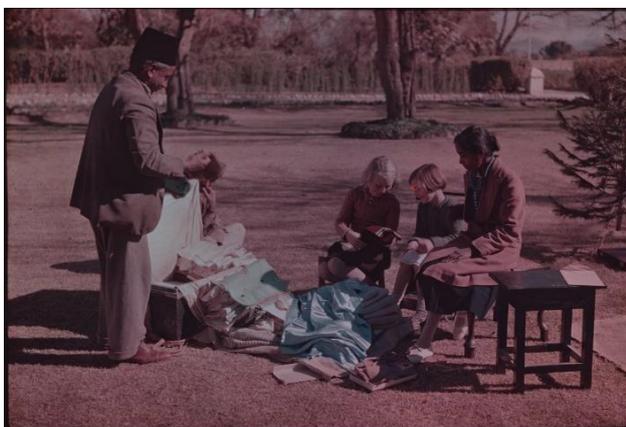 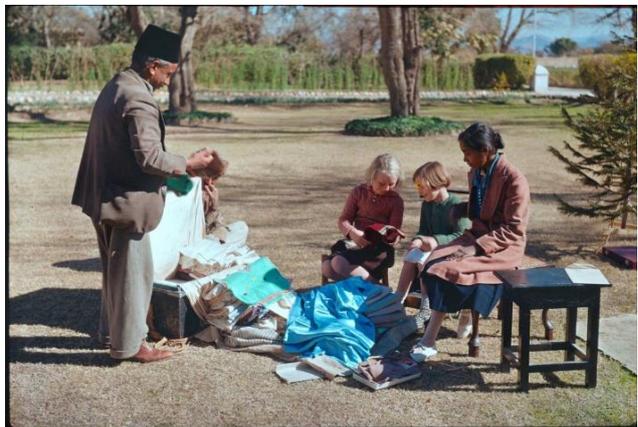

Fig. 7 - Probably J.B.S. Thubron and/or his wife Diana, Dufay color transparency, ca. 1930s, 6×9cm. Color calibrated scan (left) and digital color reconstruction based on infrared scan (right). Scanned at 4000PPI using Nikon Coolscan 9000ED.

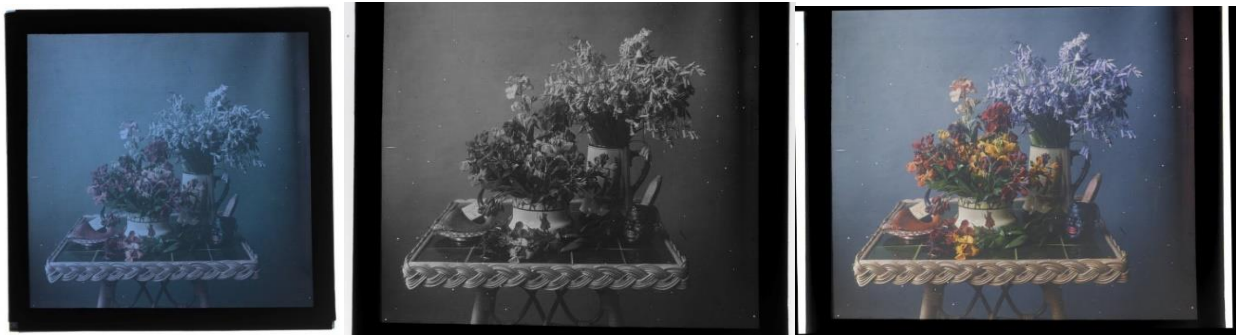

Fig. 8 - Unknown photographer, Paget lantern slide 8×8cm, ca 1921. Color calibrated scan (left), infrared scan (middle) and digital color reconstruction (right). Scanned at 4000PPI using DT Atom Rainbow MSI.

**Conclusion**

Additive color screen processes combine a viewing screen (which is not very physically stable) with a black and white transparency (that is very stable). Even if the pictures appear faded, with special image processing it is possible to digitally reconstruct the original colors with a high degree of precision. This processing stresses some qualities of the digital scan (resolution, sharpness, geometric precision, and dynamic range which exceed the needs for digitizing regular black and white photographs). These requirements, however, can be practically achieved with existing scanners.

It is our hope and expectation that this paper, along with the software we will be publishing alongside it, will help unlock the beauty and value of the original color of these collections by lifting the veil imposed on them by time.

We would like to thank to Bertrand Lavédrine for remaks which significantly improved this paper.

*Colour Photography and Film:*
*sharing knowledge of analysis, preservation, conservation, migration of analogue and digital materials*